# Mg/Si MINERALOGICAL RATIO OF LOW-MASS PLANET HOSTS. CORRECTION FOR THE NLTE EFFECTS

V. Adibekyan[1], H.M. Gonçalves da Silva[2], S.G. Sousa[1], N.C. Santos[1,2],
E. Delgado Mena[1], A.A. Hakobyan[3]

Mg/Si and Fe/Si ratios are important parameters that control the composition of rocky planets. In this work we applied non-LTE correction to the Mg and Si abundances of stars with and without planets to confirm/infirm our previous findings that [Mg/Si] atmospheric abundance is systematically higher for Super-Earth/Neptune-mass planet hosts than stars without planets. Our results show that the small differences of stellar parameters observed in these two groups of stars are not responsible for the already reported difference in the [Mg/Si] ratio. Thus, the high [Mg/Si] ratio of Neptunian hosts is probably related to the formation efficiency of these planets in such environments.

Key words: planetary systems: abundances: metallicity: Galaxy

## 1. Introduction

It is now a solid observational fact that stellar overall metallicity plays an important role on the formation of giant planets [1-4]. It seems also that the role of metallicity on the formation of low-mass planets if exist, is not strong [5-7]. These observational results are supported by the core-accretion model [8,9] and the tidal-downsizing model of Sergey Nayakshin [10]. Interestingly, however, some very new results show that indeed low-mass planets at very short orbits tend to appear around stars with super-solar metallicities [11] and their maximal mass increases with metallicity [12]. Recent works on star-planet connection, besides planet formation, also clearly demonstrated the importance of metallicity on the orbital architecture of planets [13-16].

After the detection of planet-metallicity correlation, several studies also tried to search for differences in chemical abundances of individual elements between stars hosting planets and stars that are known not to host a planet [17-24]. Although most of these studies observed no significant differences, some authors reported a clear overabundance in α-elements of stars hosting both high-mass [25-27] and low-mass planets [28] when compared to the stars with no planetary companion. Knowledge of precise abundances of individual heavy elements and elemental ratios is very important since they can control the structure and composition of terrestrial planets [21, 29-32]. In particular, [Mg/Si] and Fe/Si ratios were proposed to allow to constrain the internal structure of rocky-planets [32]. These theoretical models were recently successfully tested on three terrestrial planets by [33].

Motivated by these recent works [31,32], Adibekyan et al [34] studied the [Mg/Si] abundance ratio of stars hosting planets of different masses. The authors found that low-mass planet hosts show a high [Mg/Si] abundance ratio when compared to the field stars without detected planetary companion [34]. Together with the difference in [Mg/Si] ratio, [34] also found that the hosts of low-mass planets are slightly cooler than their counterparts without planets. While the authors warned the reader about the possible non local thermodynamic effects (non-LTE) on the [Mg/Si] abundance ratio, [34] concluded that the small differences in the stellar parameters are not expected to be responsible for the significant differences in [Mg/Si] ratio between Neptunian hosts and non-host stars.

In this paper we used the MPIA (Max Planck Institute for Astronomy) hosted web server for non-LTE stellar spectroscopy [35,36] to estimate the impact of the non-LTE corrections on our the aforementioned findings [34]. This work is organized as follows. In Section 2 we present the sample and in Section 3 we describe how the non-LTE corrections are performed. In Section 4 we present our results and summarize them in Section 5.

## 2. The sample

Our sample is taken from [33]. From an initial sample of 1111 FGK-type stars [24] we selected 587 dwarfs (log g ≥ 4 dex), that have effective temperatures within 500 K from that of the Sun ($T_\odot$ = 5777 K) and errors on Mg, Si, and Fe abundances smaller than 0.2 dex. These criteria allow us to select stars with the most precise abundance and atmospheric parameter determinations [24,37]. Moreover, we only considered stars that have metallicities from -0.6 to 0.4 dex, which are the [Fe/H] limits of planet hosting stars in the sample. This sample consists of 489 stars without detected planetary companions, 19 stars hosting super-Earths or Neptune-like planets (the most massive planet in the system has $M_p < 30\ M_\oplus$) and 79 Jovian hosts ($M_p \geq 30\ M_\oplus$).

## 3. NLTE correction

Several works in the literature have already tried to quantify the non-LTE effects in Mg and Si, and their dependence on the stellar parameters [35,36,38-43]. They show that the non-LTE effects are usually stronger for evolved, hot, and metal-poor stars.

MPIA (Max Planck Institute for Astronomy) hosted web server provides non-LTE corrections for a large list of lines of several chemical elements. This database also contains non-LTE corrections for all the three Mg spectral lines (4730.030, 5711.07, and 6319.24 ÅÅ) used in [34] and four (5645.66, 5684.52, 5772.15, 5948.54 ÅÅ) out of 16 Si lines that were used to derive average Si abundances in [34]. The non-LTE corrections for the mentioned lines were extracted from this database. The amplitudes of the corrections for these lines are quite different. For example, the correction for 5772.15Å Si line is essentially zero for more than 90% of the stars, and the maximum correction for this line is -0.01 dex for our sample stars. In contrast, the correction for 5711.07Å Mg line lies between -0.03 and 0.03 dex, and is negligible only for stars that have effective temperature very close to the Sun (5600 to 5800K).

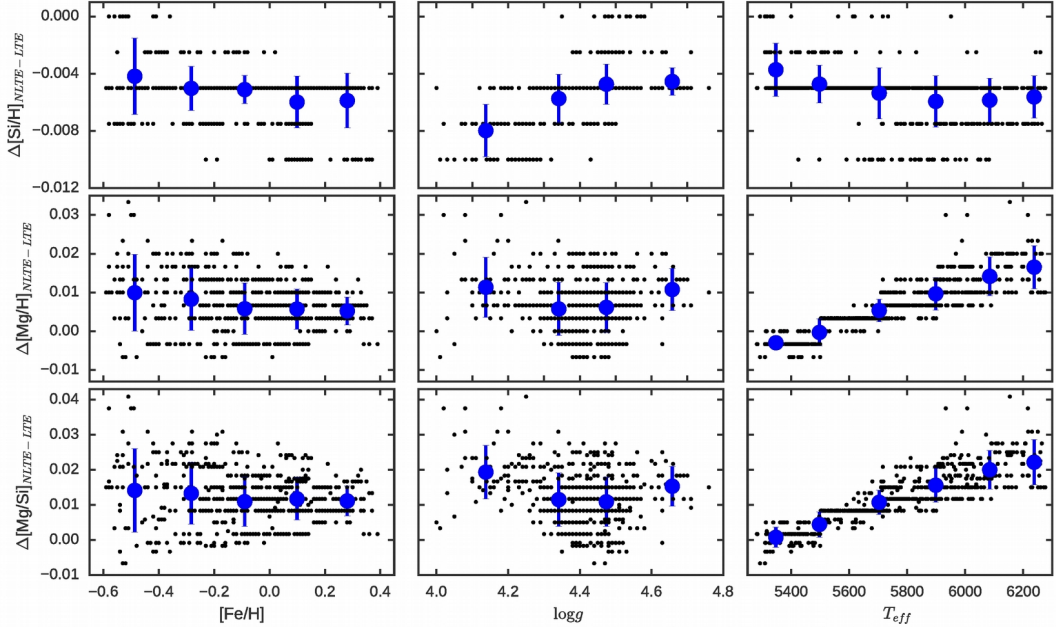

Fig. 1. Average non-LTE correction of [Mg/H], [Si/H], and [Mg/Si] against stellar parameters for the stars in our sample (black dots). The average value for different bins of stellar parameters are shown in blue filled circles. The bin size is 0.2 dex for [Fe/H] and log g, and is 200K for $T_{eff}$. The error bar represents the standard deviation (dispersion) for each bin.

To apply the non-LTE correction for our sample stars we used the mean non-LTE correction for all the available lines (three for Mg and four for Si). The average correction and standard deviation of [Mg/H], [Si/H], and [Mg/Si] abundance corrections against stellar parameters are presented in Fig. 1. The figure shows that the correction is small and always negative (from ~ -0.004 to ~0.012 dex) for Si. It mostly depends on effective temperature and surface gravity: hotter and evolved stars show the largest deviations from LTE. The absence of the correlation with metallicity is probably due to a wide range of $T_{eff}$ and log g at a given metallicity and inter-correlation between stellar parameters. Fig. 1 also shows that the correction for Mg is stronger than for Si. The amplitude of the correction is ~0.04 dex: from ~-0.01 to ~0.03 dex. This correction mostly depends on effective temperature being negative for stars cooler than our Sun and positive for hotter stars. One can also notice that the non-LTE corrections both for Mg and Si are tight for cooler stars ($T_{eff}$ < ~5500 K), and get more dispersed for hotter stars. This is because almost all the cool stars have log g of nearly 4.4 dex, while hotter stars have a wider range of log g values. Finally, in Fig. 1 one can see that the non-LTE correction for [Mg/Si] ratio is almost zero for cool stars and reaches up to 0.03 dex for the hottest stars.

## 4. Results

After applying the non-LTE corrections for [Mg/Si] ratio we proceed in the same way as in [34]. In Fig. 2 (left panel) we plot [Mg/Si] against metallicity for stars without planetary mass companions and stars that are hosting Neptunes and Jupiters. The top right panel of the same figure shows the cumulative distributions of [Mg/Si] ratio for the three groups of stars. The figure shows that even after the corrections [Mg/Si] ratio is higher for low-mass planet hosts and lower for high-mass planet hosts when compared to their non-host counterparts. To quantify these differences we applied the two-sample Kolmogorov-Smirnov (KS) and the Anderson-Darling (AD) tests. The results of the tests (Table 1) show that the difference is significant (P ≤ 0.05) only when comparing Jovian-hosts with non-hosts. We remind the reader that if the non-LTE corrections are not applied, the KS and AD tests predict significant differences in the [Mg/Si] distributions between Neptunian-hosts and non-host stars (see Table 1 of [34]).

However, as discussed in our previous work [34] and as can be clearly seen in Fig. 1, the [Mg/Si] ratio depends on the metallicity. This dependence reflects the Galactic chemical evolution (GCE). It is also known that stars without planets and stars with planets of different masses show different metallicity distributions (see the introduction). These correlations can obviously affect our comparison of [Mg/Si] ratio of the three groups. To remove the trend of GCE from the [Mg/Si] ratio, as in [34], we performed a linear fit to our data points (all stars with and without detected planets) and then subtracted the fit. We applied the KS and AD tests to the subtracted data again, and found that the difference in [Mg/Si] between Jovian hosts and the single stars disappears (see Table 1). This result is in agreement with [34] and suggests that the difference observed in the original dataset was just a reflection of GCE and the shift of giant-planet hosts towards high metallicities.

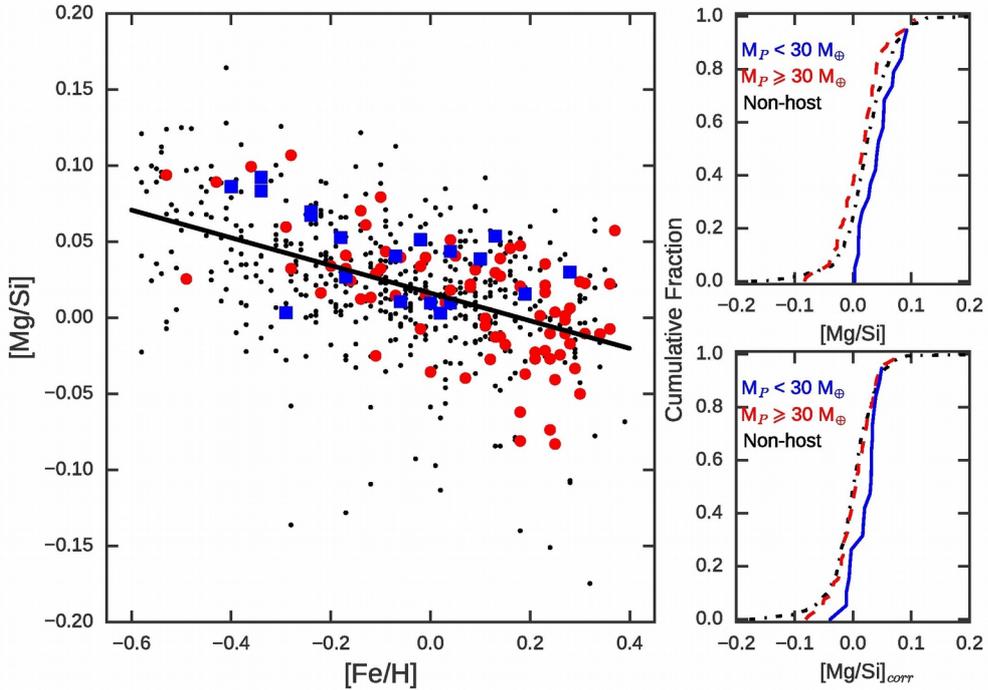

**Fig. 2.** Left: [Mg/Si] against [Fe/H] for stars without detected planets (black dots), stars hosting low-mass planets (blue squares), and stars hosting high-mass planets (red filled circles). The black solid line shows the linear fit to all data points. Cumulative distribution of [Mg/Si] (top right) and [Mg/Si]$_{cor}$ after correcting for the GCE (bottom right) for stars with giant planets (blue solid line), low-mass planets (red dashed line), and stars without detected planets (black dotted-dashed line).

*Table 1*
KS AND AD PROBABILITIES THAT PLANET HOSTS AND STARS WITHOUT DETECTED PLANETS COME FROM THE SAME PARENT DISTRIBUTION

|          |                      | [Mg/Si] in LTE | | [Mg/Si] in NLTE | |
|----------|----------------------|----------|---------------|---------|----------------|
|          |                      | [Mg/Si]  | [Mg/Si]$_{corr}$ | [Mg/Si] | [Mg/Si]$_{corr}$ |
| $P_{KS}$ | Neptune – Non-host   | **0.037** | **0.009**     | 0.097   | **0.005**      |
|          | Jovian – Non-host    | 0.051    | 0.736         | **0.050** | 0.872        |
| $P_{AD}$ | Neptune – Non-host   | **0.020** | **0.003**     | 0.054   | **0.010**      |
|          | Jovian – Non-host    | **0.017** | 0.841         | **0.015** | 0.729        |

The statistically significant differences between the distributions are in boldface.

In contrast to Jovian hosts, the difference in [Mg/Si] after correction for the GCE between Neptune hosts and non-hosts becomes significant. The applied statistical tests predict low probabilities that the two samples come from the same parent distribution (Table 1). This result confirms our previous findings that low-mass planet hosts in average have higher [Mg/Si] ratio than stars without planets. To evaluate the statistical significance of the observed difference in [Mg/Si]$_{cor}$, we followed [34] and performed two simple Monte Carlo tests. The tests show that there is a low chance (~ 0.5%) of obtaining the observed difference by chance.

## 5. Summary

We applied non-LTE corrections to the Mg and Si abundances of 587 dwarf stars from which 19 are known to host a low-mass planet ($M_p < 30$ $M_⊕$) and 79 are hosting Jupiters ($M_p \geq 30$ $M_⊕$). With these non-LTE corrected abundances we confirmed our previous results that [Mg/Si] mineralogical ratio is slightly, but statistically significantly, higher for stars that are hosting low-mass planets when comparing with stars without detected planetary companion. The systematic overabundance of Mg relative to Si ([Mg/Si] ratio) is always positive for the Neptunian hosts) probably means that this ratio plays an important role in the formation of low-mass planets. It is worth to remind that this ratio is expected to play a very determining role on the structure and composition of terrestrial planets [30-32].

**Acknowledgements**
V.A., E.D.M., N.C.S. and S.G.S. acknowledge the support from Fundação para a Ciência e Tecnologia (FCT) through national funds and from FEDER through COMPETE2020 by the following grants UID/FIS/04434/2013 & POCI-01-0145-FEDER-007672, PTDC/FIS-AST/7073/2014 & POCI-01-0145-FEDER-016880 and PTDC/FIS-AST/1526/2014 & POCI-01-0145-FEDER-016886. V.A., N.C.S. and S.G.S. also acknowledge the support from FCT through Investigador FCT contracts IF/00650/2015, IF/00169/2012/CP0150/CT0002 and IF/00028/2014/CP1215/CT0002; and E.D.M. acknowledges the support by the fellowship SFRH/BPD/76606/2011 funded by FCT (Portugal) and POPH/FSE (EC). V.A. thanks Pedro Figueira for interesting comments.

[1] Instituto de Astrofísica e Ciências do Espaço, Universidade do Porto, CAUP, Rua das Estrelas, 4150-762 Porto, Portugal, e-mail: vadibekyan@astro.up.pt
[2] Departamento de Física e Astronomia da Faculdade de Ciências da Universidade do Porto, Portugal
[3] Byurakan Astrophysical Observatory, Armenia